\documentclass[prl,aps,reprint,letterpaper,nofootinbib,superscriptaddress,floatfix]{revtex4-2}
\usepackage{times}
\usepackage{graphicx}
\usepackage{epsfig}
\usepackage{amsfonts}
\usepackage{amsmath}
\usepackage{amssymb}
\usepackage{dsfont}
\usepackage{amsthm}
\usepackage{color,xcolor}
\usepackage[normalem]{ulem}
\usepackage[colorlinks=true,linkcolor=blue,citecolor=blue,urlcolor=blue,breaklinks]{hyperref}
\usepackage{comment}
\usepackage{braket}
\usepackage{physics}
\usepackage[breaklinks]{hyperref}
\usepackage{bbm}

\usepackage[T1]{fontenc}
\usepackage{braket}

\begin{document}
	
\title{Noise-Resilient Quantum Random Access Codes}

\author{H. S. Karthik$^{\ddagger}$}
\email{hsk1729@gmail.com}
\affiliation{International Centre for Theory of Quantum Technologies, University of Gda{\'n}sk, 80-308 Gda\'nsk, Poland.}

\author{S. G\'omez$^{\ddagger}$}
\affiliation{Departamento de Física, Universidad de Concepción, Casilla 160‑C, Concepción, Chile.}
\affiliation{Millennium Institute for Research in Optics, Universidad de Concepción, Casilla 160‑C, Concepción, Chile.}

\def\thefootnote{$\ddagger$}\footnotetext{These authors contributed equally to this work}\def\thefootnote{\arabic{footnote}}

\author{F. M. Quinteros}
\affiliation{Departamento de Física, Universidad de Concepción, Casilla 160‑C, Concepción, Chile.}
\affiliation{Millennium Institute for Research in Optics, Universidad de Concepción, Casilla 160‑C, Concepción, Chile.}

\author{Akshata Shenoy H.}
\affiliation{International Centre for Theory of Quantum Technologies, University of Gda{\'n}sk, 80-308 Gda\'nsk, Poland.}

\author{M. Paw{\l}owski}
\affiliation{International Centre for Theory of Quantum Technologies, University of Gda{\'n}sk, 80-308 Gda\'nsk, Poland.}

\author{S. P. Walborn}
\affiliation{Departamento de Física, Universidad de Concepción, Casilla 160‑C, Concepción, Chile.}
\affiliation{Millennium Institute for Research in Optics, Universidad de Concepción, Casilla 160‑C, Concepción, Chile.}

\author{G. Lima}
\affiliation{Departamento de Física, Universidad de Concepción, Casilla 160‑C, Concepción, Chile.}
\affiliation{Millennium Institute for Research in Optics, Universidad de Concepción, Casilla 160‑C, Concepción, Chile.}

\author{E. S. G\'omez}
\email{estesepulveda@udec.cl}
\affiliation{Departamento de Física, Universidad de Concepción, Casilla 160‑C, Concepción, Chile.}
\affiliation{Millennium Institute for Research in Optics, Universidad de Concepción, Casilla 160‑C, Concepción, Chile.}

\begin{abstract}
A $n^d \xrightarrow{p} 1$ Quantum Random Access Code (QRAC) is a communication task where Alice encodes $n$ classical bits into quantum states of dimension $d$ and transmits them to Bob, who performs appropriate measurements to recover the required bit with probability $p$. In the presence of a noisy environment, the performance of a QRAC is degraded, losing the advantage over classical strategies. We propose a practical technique that enables noise tolerance in such scenarios, recovering the quantum advantage in retrieving the required bit. We perform a photonic implementation of a $2^2 \xrightarrow {\text{p}} 1$ QRAC using polarization-encoded qubits under an amplitude-damping channel, where simple operations allow for noise robustness showing the revival of the quantum advantage when the noisy channel degrades the performance of the QRAC. This revival can be observed by violating a suitable dimension witness, which is closely related to the average success probability of the QRAC. This technique can be extended to other applications in the so-called prepare-and-measure scenario, enhancing the semi-device-independent protocol implementations.
\end{abstract}
\maketitle

\textit{Introduction.}-- Quantum communication is quite unique as compared to its classical counterpart due to the fundamental differences in their physical principles \cite{GRT+02,NT07}. Quintessential quantum features like quantum superposition, entanglement, and no-cloning theorem form the basis of security and privacy in any quantum communication protocol \cite{BB84,E91}. Amongst them, QRAC is a well-known communication protocol where a sender, Alice, encodes a string of classical bits onto a quantum state and sends it to a receiver, Bob, who needs to extract the required bit by performing appropriate measurement \cite{ANT+99,ANT+02}. The performance of this protocol is measured in terms of Bob's average success probability (ASP) of extracting the value of the desired bit. For instance, in a $2^{d=2} \xrightarrow {\text{p}} 1$ QRAC protocol, Alice encodes two bits ($a_0,a_1)$ onto quantum states of a single qubit ($d=2$). The prepared state is transmitted to Bob, who extracts the bit value by performing an optimal measurement called the \textit{strategy}. The ASP for this QRAC is $p=0.85$, which is greater than the case for the optimal classical strategy ($p_c=0.75$), exhibiting a quantum advantage \cite{ALM+09}. The classical strategy consists of Alice encoding and sending the same bit always--say it is $a_0$.  If Bob is interested in $a_0$, he is always correct, and if he is interested in $a_1$, he simply has to guess which gives the success probability as $\frac{1}{2}+\frac{1}{2}.\frac{1}{2}=\frac{3}{4}$. It has been demonstrated that the performance of a QRAC protocol can be further enhanced by using entanglement-assistance \cite{PZ10} or using $d$-level quantum systems \cite{THB+15}.

To understand the overall performance of any quantum communication protocol, the assumptions used to derive its security parameters are crucial. Given that a $2^{2} \xrightarrow {\text{p}} 1$ QRAC is represented as a \textit{prepare-and-measure protocol}, one finds it apt to consider such performance in the \textit{semi-device-independent (SDI)} paradigm with the following main assumptions: (i) devices used for state preparation and measurement are uncharacterized, and (ii) the dimension of the Hilbert space of the prepared quantum system is bounded. The dimension of a system is an important resource in quantum theory. In SDI scenarios, the emphasis is only on analyzing the data collected in an experiment to characterize the dimension of a system independent of any theory. In this regard, a suitably constructed dimension witness (DW) allows one to certify the given Hilbert space dimension by placing an upper bound on it \cite{GBH+10,BQB14,ABC+12,HGM+12,ABP+14}. In a given scenario involving a system of dimension $d$, the saturation of the upper bound corresponds to the presence of a classical or quantum system of dimension $d$. On the other hand, given a classical and a quantum system of the same dimension, these witnesses can be violated by the quantum system, distinguishing it as a potentially more useful resource in comparison to its classical counterpart (say for realizing some quantum communication protocol). Thus, the classical system needs to be augmented with more dimension to reproduce the observed measurement statistics. It has been shown that the problem of constructing a dimension witness is related to the designing of QRACs \cite{WCD08}. Thus, by utilizing the connection between DWs, QRACs, and quantum key distribution protocols (QKD), the security of any prepare-and-measure protocol in the SDI scenario can be established under individual attacks by checking either the DW or the ASP of a QRAC protocol \cite{PB11, CRV+18}. 

Some other important applications of QRACs include SDI random number expansion \cite{WZY+11, MP22}, randomness certification \cite{LPY+12}, SDI-QKD and foundational studies \cite{GHH+14, PPK+09}. Additionally, in the prepare-and-measure scenario, sequential QRACs are employed for self-testing of quantum measurements performed by an untrusted device\cite{MTB19,MBP20,FCV+20}.

It is surprising to note that all these investigations involving QRACs have been carried out in an ideal, noiseless setting. Recently, Ref. \cite{MS22} considered the effect of different kinds of noise on a QRAC protocol and reported that the noise in the channel causes degradation in a manner that even a classical RAC could outperform it. They also attempt to mitigate these losses by considering semi-definite programming techniques in scenarios when the channel noise is known. 

In this letter, we consider a conceptually different approach where we address this issue from a practical operational perspective, aiding experimental implementations in a more convenient way. To this end, we experimentally consider the noise channel to be an amplitude damping channel (ADC), which models the effects of energy dissipation or loss of energy of a quantum state into the environment. We present a method to \textit{activate} a QRAC protocol and restore its overall performance using suitable stochastic operations. The main conceptual novelty of this work is in the application of these simple operations, which aid in overcoming the effects of such noise and can be extended to other applications in the so-called prepare and measure scenario, enhancing the semi-device-independent protocol implementations.

\begin{figure}[!t]
\includegraphics[width=0.45\textwidth]{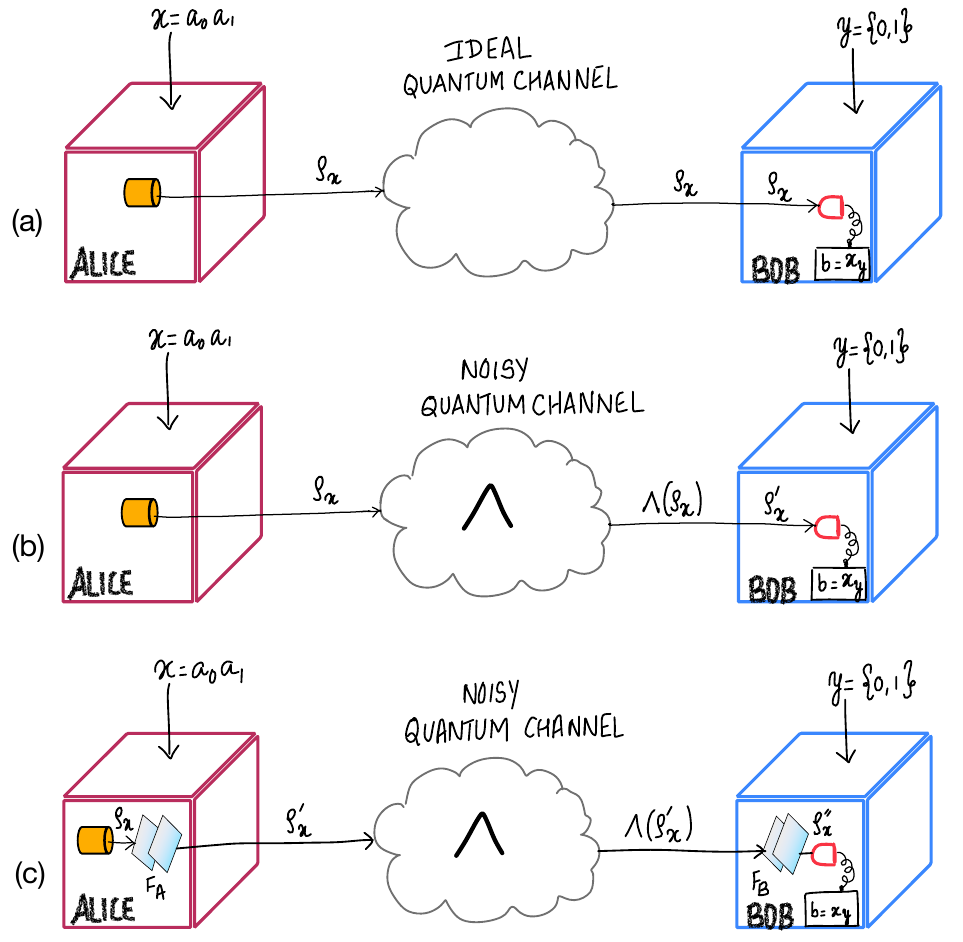}
\caption{A schematic representation of (a) QRACs in the ideal SDI scenario when Alice prepares and transmits states $\rho_x$ to Bob. Quantum advantage is observed when Bob recovers the required subset of bits with an ASP $P_b > P_b ^{class} = 0.75$ by measuring $M_y$. Further, violation of the DW by the encoded states certifies that they are quantum states of dimension $2$, ensuring semi-device independence. (b) QRACs in the SDI scenario when Alice's states are transmitted over a noisy quantum channel $\Lambda$. After transmission along the channel, the state $\rho'_x=\Lambda(\rho_x)$ reaches Bob. (c) QRACs in the noisy SDI scenario with activation. A filter $F_A$ is applied at Alice's end before transmitting the encoded state, $\rho_x$, rendering the new state $\rho'_x$. After transmission along the channel, the state $\Lambda(\rho'_x)$ is again filtered by Bob using $F_B$ before performing measurements $M_{y}$ on the final state $\rho''_x$.}
\label{general}
\end{figure}

\textit{QRACs.}-- A $2^2 \xrightarrow {\text{p}} 1$  QRAC protocol involves Alice encoding a uniformly random two-bit string $x=a_0a_1 \in\{0,1\}^2$ onto a two-dimensional quantum state $\rho_x$. Given $x$, Alice prepares and transmits the state to Bob. He wants to recover a subset of the encoded bit string $x$ by performing a projective measurement $M^{b}_y$ defined by the random classical bit $y\in\{0,1\}$ with outcomes $b \in \{0,1\}$. The ASP of Bob in extracting the required bit can be written as
\begin{equation}
P_b = \frac{1}{8} \sum_{x} \mathrm{Tr}\left[\rho_x (M_0+M_1)\right], \label{asp}
\end{equation}  
where Bob's measurements are defined by the operators $M_j= \frac{\mathbbm{1}+ n_j \Vec{\sigma}\cdot\Vec{n_j}}{2}$, with $\vec{\sigma}=(\sigma_x,\sigma_y,\sigma_z)$, the outcomes $n_j = \pm1$, $j=0,1$ and $\Vec{n_j}$ are vectors in the Bloch sphere. A QRAC is successfully implemented if the ASP of extracting each of the bits transmitted by Alice exceeds the optimal classical ASP $P_b ^{class}=0.75$.

\textit{Semi-device independence and dimension witness.}-- In the semi-device independence scenario, no assumptions are made about the preparation and measurement of quantum states. In contrast, an upper limit for the dimension of the states is assumed. \cite{PB11}. Such a scenario is depicted in Fig.~\ref{general}(a). Alice prepares states $\rho_x$ with the assumption that everything except the dimension of the Hilbert space is uncharacterized. Now, Bob can apply projective measurements ${M^b_y}$ with outcomes $b\in\{0,1\}$ to obtain the conditional probability distribution $P(b|x,y)=\mathrm{Tr}(\rho_{x}M^b_y)$. This can be used to construct a DW of the form
\begin{eqnarray}
W &=& E_{00,0}\,+\, E_{00,1}\,+\, E_{01,0}\,-\, E_{01,1} \nonumber\\ 
    &-& E_{10,0}\,+\, E_{10,1}\,-\, E_{11,0}\,-\,E_{11,1} \,\le\, 2,
\label{witness}
\end{eqnarray}
which is a linear inequality of the conditional probability distribution $E_{a_0a_1,y} = P(b=0|a_0a_1,y) = \mathrm{Tr}(\rho_{a_0a_1} M^{b=0}_y)$. Here, $\rho_{x=a_0a_1}$ is an encoded state on which the measurement $M^{b=0}_y$ acts. Without loss of generality, only the projector corresponding to the outcome value $b=0$ is considered. The upper bound of the $W$ corresponds to the optimal capacity of a classical system of dimension $2$ in conveying $1$ bit of communication. On the other hand, the inequality is violated by qubits with a maximum value $2\sqrt{2}$, thereby establishing their resourceful nature in transferring enhanced communication $>1$ bit. Moreover, a generalization for any arbitrary dimension $d$ is possible \cite{GBH+10}. Thus, DWs aid in characterizing any arbitrary system of dimension $d$ under consideration as classical or quantum, in a \textit{semi-device independent} way, without any characterization of the experimental setup used to perform these tests.

\begin{figure*}[th!]
\centering
\includegraphics[width=17cm]{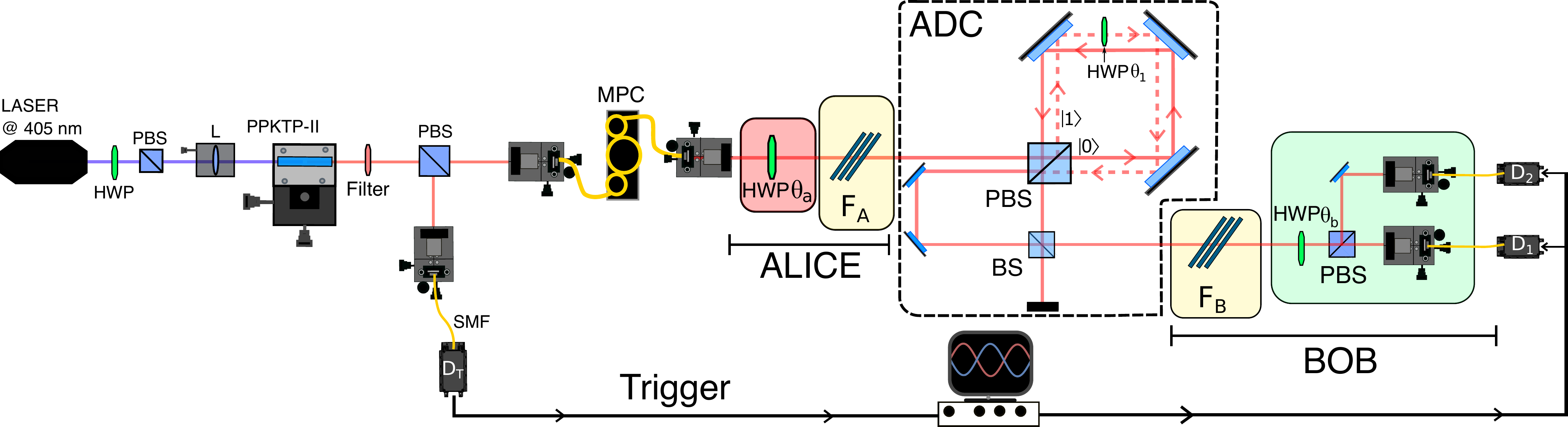}
\caption{Photonic implementation used to study the behavior of the SDI QRAC protocol in a noisy scenario. A heralded photon source based on the SPDC process is implemented to produce single photons. Alice's setup includes an HWP$(\theta_{a})$ that is used to prepare four states encoded in the polarization degree of freedom: $\rho_{00}$, $\rho_{11}$, $\rho_{01}$, and $\rho_{10}$. Additionally, a filter denoted $F_{A}$ is applied before sending the state. Subsequently, a two-path Sagnac interferometer implements the ADC, where $\gamma$ is controlled by $\theta_{1}$. After transmitting the state through the channel, Bob applies $F_{B}$, and finally, two measurements are performed by an HWP, a PBS, and two photon detectors. The stochastic filters $F_{A}$ and $F_{B}$, are implemented using thin glass plates placed at the Brewster angles.}
\label{Fig1}
\end{figure*}

Rewriting Eq.(\ref{asp}) using the above parameters \cite{PB11}, we obtain a relationship between ASP and DW in the SDI scenario as follows
\begin{equation}
P_b = \frac{1}{8} \sum_{x,y} P(b=x_y|a_0a_1,y) = \frac{W+4}{8}
\label{aspw}.
\end{equation} 
From the above equation, we can infer that, if $W$ is violated given a physical system of dimension $d=2$, then a \textit{quantum} system of $d=2$ was prepared. Correspondingly, a QRAC protocol with ASP $P_b > 0.75$ can be realized. If $W$ is not violated, we cannot certify whether a quantum system or a classical system of $d=2$ was prepared. Thus, DWs are useful tools in certifying the quantumness of a system given the dimension $d$ and in the case of $d=2$ also warrants the realization of a $2^2 \xrightarrow{\text{P}} 1$ QRAC protocol in the SDI scenario.

\textit{Noise-adapted QRACs.}-- We now consider the SDI-QRAC protocol's performance under a noisy channel's action. Specifically, we will consider the ADC denoted by $\Lambda$, where it acts on the encoded states being transmitted (see Fig.~\ref{general}(b)). After the transmission, Bob receives each encoded state affected by the noise. ADC models the process of spontaneous emission of a quantum system dissipating energy into the environment. An ADC, parameterized by $\gamma$, is represented in terms of Kraus operators $K_0(\gamma)$ and $K_1(\gamma)$, such that
\begin{eqnarray}
\label{ampdamp}
K_0(\gamma) &=& |0\rangle\langle0| + \sqrt{1-\gamma} |1\rangle\langle1| \\\nonumber
K_1(\gamma) &=& \sqrt{\gamma} |0\rangle\langle1|,
\end{eqnarray} 
where $K_i(\gamma) \ge 0$, $\sum_i K_i^{\dagger}(\gamma) K_i(\gamma) = \mathbbm{1}$, and $\gamma \in [0,1]$. Thus, Alice's states after the transformation due to the amplitude damping can be written as
\begin{equation}
\Lambda(\rho_x)=\sum_{i} K_i(\gamma) \, \rho_x \, K_i^{\dagger}(\gamma).
\label{stateev}
\end{equation} 
Therefore, the corresponding expressions for the DW and ASP considering the ADC are
\begin{eqnarray}
\label{eq:aampdampw}
W(\gamma) &=& \sqrt{2}\left(1+\sqrt{1-\gamma}-\gamma\right) \\\nonumber
P_b(\gamma) &=& \frac{W(\gamma) +4}{8}.
\end{eqnarray} 
In the absence of noise ($\gamma =0$), the maximum achievable ASP is $P_b \approx 0.85$, corresponding to the maximal violation of $W$ given by $W=2\sqrt{2}$. Then, we define $\gamma_c$ as the critical amplitude damping parameter beyond which the SDI-QRAC protocol is unsuccessful. From Eq. (\ref{eq:aampdampw}), it is seen that the QRAC no longer surpasses the optimal classical ASP of 3/4 at $\gamma_c \approx 0.375$. For $\gamma < \gamma_{c}$, a violation of $W(\gamma)$ proves the realization of an SDI-QRAC scheme with $P_b > 3/4$.

It has been recently identified that it is possible to overcome the effects of an amplitude-damping noise upon the application of certain stochastic operations at appropriate points along the channel \cite{HH01}. In this sense, \textit{activation} is restoring the quantum advantage of an SDI-QRAC protocol under the action of noise, recovering the violations of the DW by applying these stochastic operations. In our case, considering the prepare and measure scenario, the stochastic operations are applied by Alice at the input of the ADC and by Bob before his measurement apparatus respectively (see Fig.~\ref{general}(c)) \cite{KWS+21}. These operations act as filters on the noisy encoded state $\Lambda(\rho_x)$. For the ADC, these filtering operations are given as 
\begin{eqnarray}
F_A &=& |0\rangle\langle0| + \sqrt{(1-f)} |1\rangle\langle1| \\ \nonumber
F_B &=& \sqrt{(1-f)} |0\rangle\langle0| + |1\rangle\langle1|,
\label{eq:norif}
\end{eqnarray} where $f\in[0,1]$ is the filter parameter. Therefore, a noise-adapted SDI-QRAC protocol can now be formulated as follows: 
\begin{enumerate} 
\item The state $\rho_x\in\{\rho_{00},\rho_{01},\rho_{10},\rho_{11}\}$ is prepared by Alice, who then applies a filtering operation $F_A$
before transmission across an ADC $\Lambda$.
\begin{equation}
\rho_x \xrightarrow {F_A} \rho'_x = \frac{F_A\rho_x F_A^\dagger}{\mathrm{Tr}(F_A\rho_x F_A^\dagger)}.
\end{equation}

\item The amplitude damped state $\Lambda(\rho'_x)$ is received by Bob. Bob applies another filtering operation $F_B$
on this state to obtain
\begin{equation}
\Lambda(\rho'_x) \xrightarrow {F_B} \rho''_x=\frac{F_B\Lambda(\rho'_x) F_B^\dagger}{\mathrm{Tr}(F_B \Lambda(\rho'_x) F_B^\dagger)}.
\end{equation}

\item Bob performs the measurement $M_y$ on the received state $\rho''_x$ to extract the required bit.
\end{enumerate}

Therefore, the probability distribution of the measurement outcome $b$ at Bob's end is given as, 
\begin{equation}
\label{prob}
p(b|y,x) = \mathrm{Tr}\left[\Pi_y^b~ \left(\frac{F_B \circ \Lambda \circ F_A (\rho_x)}{N(x)}\right)\right],
\end{equation}
where
\begin{equation*}
N(x) = \mathrm{Tr}\left[F_B \left(\sum_i K_i(\gamma) (F_A~\rho_x~F_A^\dagger) K_i(\gamma)^\dagger\right) F_B^\dagger\right],
\end{equation*} is the normalization constant
and $\Pi_y^b$ is the projector associated with the measurement $y$ and the outcome $b$.
\\

\textit{Experimental demonstration.}-- Our noise-adapted implementation of the QRAC is depicted in Fig.~\ref{Fig1}. The experiment consists of (i) a sender (Alice) who encodes four states into the polarization state of single photons, (ii) a receiver (Bob) who performs projective measurements on the photons, (iii) a noisy channel through which the polarization qubits propagate from Alice to Bob, and (iv) a set of filter operations $F_A$ and $F_B$ that Alice and Bob will use to attempt to improve the performance of the protocol in the presence of noise. 

Single photons are produced using a heralded single-photon source based on the spontaneous parametric down-conversion (SPDC) process \cite{SupMat}. During the SPDC process, a pair of photons with orthogonal polarization at $810$ nm are produced. A polarizing beam splitter (PBS) is placed to separate the generated photons by polarization. One photon is sent directly to a single photon detector $D_T$, which acts as a trigger, while the other photon is sent to the Alice stage to initialize the protocol. Alice uses a half-wave plate (HWP) to prepare polarization qubits of the form $\ket{\psi}=\cos(2\theta_{a})\ket{H}+\sin(2\theta_{a})\ket{V}$, where $\theta_{a}$ is the angle of the HWP concerning its fast axis. The four states prepared by Alice for use in the protocol are 
\begin{align}
    \begin{split}
    &\rho_{00} = \ket{H}\bra{H}, \quad \rho_{01} = \ket{-}\bra{-},\\
    &\rho_{10} = \ket{+}\bra{+}, \quad \rho_{11} = \ket{V}\bra{V},
    \end{split}
    \label{states}
\end{align}
where $\ket{\pm} = \left(\ket{H}\pm \ket{V}\right)/\sqrt{2}$ are the diagonal and anti-diagonal polarization states, respectively. 

Bob's optimal measurements for extracting the required bit in the SDI-QRAC protocol are given as, 
\begin{eqnarray}
M_{y=0} &=& \frac{1}{\sqrt{2}} (\sigma_z - \sigma_x), \\ \nonumber
M_{y=1} &=& \frac{1}{\sqrt{2}} (\sigma_z + \sigma_x),
\end{eqnarray}
where $\sigma_z=\ket{H}\bra{H}-\ket{V}\bra{V}$ and $\sigma_x=\ket{+}\bra{+}-\ket{-}\bra{-}$ are the Pauli matrices. The measurements are implemented using the HWP$(\theta_{b})$ setting at $\theta_{b}=\{\frac{7\pi}{16},\frac{\pi}{16}\}$ for $y={0,1}$, respectively, and a polarizing beam splitter followed by two avalanche single photon detectors $D_{1}$ and $D_{2}$, as shown in Fig.~ \ref{Fig1}. The signals from $D_{1}$ and $D_{2}$ are processed by a coincidence counting module with calibrated delays between the output of $D_1$ ($D_2$) and Alice's $D_T$. The module has a $1$ ns coincidence window to reduce accidental counts. 

The ADC, represented as a decomposition of the Kraus operators $K_{0}(\gamma)$ and $K_{1}(\gamma)$ given in Eq. \eqref{ampdamp}, is implemented by a two-path Sagnac interferometer \cite{almeida07,salles08, PhysRevA.97.040102, PhysRevLett.117.260401, PhysRevA.94.042309}. As described in detail in Refs. \cite{almeida07,salles08}, by incoherently combining the two outputs of the interferometer, we can implement a controllable decoherence channel described by the Kraus operators
\begin{eqnarray}
\label{eq:Kraus}
    K_0 (\theta_{1}) &=& \ket{H}\bra{H} + \cos{(2\theta_{1})}\ket{V}\bra{V}, \\\nonumber
    K_1 (\theta_{1}) &=& \sin{(2\theta_{1})}\ket{H}\bra{V}.
\end{eqnarray} 
More details about the optical decoherence channel are given in the supplementary material (section B) \cite{SupMat}. Comparing these expressions with Eq. \eqref{ampdamp} we see  that $\sin{2\theta_{1}} = \sqrt{\gamma}$ and $\cos{2\theta_{1}} = \sqrt{1 - \gamma}$, allowing for control of $\gamma$ using the waveplate angle $\theta_1$.

Finally, the filter operations $F_A$ and $F_B$ (see Eq.\eqref{eq:norif}) are implemented using thin glass plates placed at the Brewster angle so that the reflected light contains only one polarization component \cite{KBSG01}. Thus, one polarization mode is partially filtered from the transmitted beam, while the other is left unchanged. The difference between $F_{A}$ and $F_{B}$ lies in the position of the principal axis of the glass plate, which are orthogonal to each other, allowing the vertical or horizontal components to be filtered.

\begin{figure}[!t]
\includegraphics[scale =0.35]{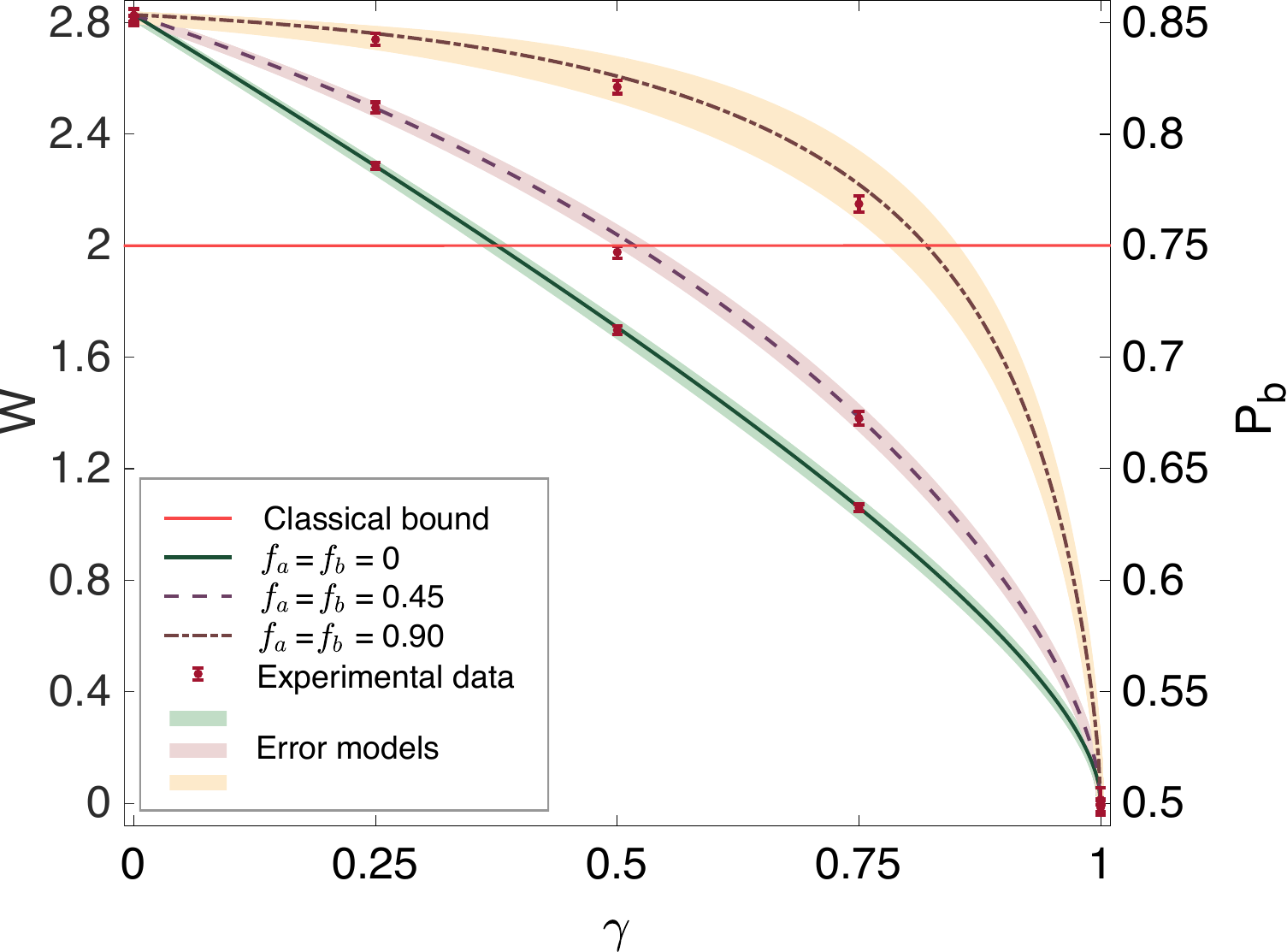}
\caption{The DW (ASP) as a function of amplitude damping parameter. The horizontal line represents the classical bound (the ASP of a classical RAC), which is $2$ (0.75). The solid, dashed, and dotted lines represent the theoretical predictions when $f_{a}=f_{b}=\{0,0.45,0.90\}$. The red dots are the experimental results. The error bars are obtained using Gaussian error propagation and considering the Poisson statistics of the recorded coincidence counts. The green, red, and yellow areas illustrate how the DW and ASP change when common experimental errors are included, such as half-wave plate settings or errors characterizing $f_{a}$ and $f_{b}$.}
\label{AvsG}
\end{figure}

\textit{Results.}--To experimentally demonstrate the activation of the quantum advantage of the SDI-QRAC protocol by applying the proposed stochastic filters, we measured the DW as a function of five different $\gamma=\{0,0.25,0.5,0.75,1\}$ and then calculated the ASP using Eq. \eqref{eq:aampdampw} (see Fig.~\ref{AvsG}). As we observed in the experimental description, it is possible to implement different values of $\gamma$ changing the angle $\theta_{1}$. For instance, when $\theta_1=0$, all photons leave the two-path Sagnac interferometer through output $\ket{0}$, indicating no amplitude attenuation in the channel, as expected. In addition, when $\theta_{1}\neq 0$, the photons start to populate the $\ket{1}$ mode, indicating an increase in noise.

The probabilities $E_{a_{0}a_{1},y}$ are calculated using the expression $E_{a_{0}a_{1},y}=\frac{cc(0|a_{0}a_{1},y)}{cc(0|a_{0}a_{1},y)+cc(1|a_{0}a_{1},y)}$. Where $cc(0|a_{0}a_{1},y)$ $\big( cc(1|a_{0}a_{1},y)\big)$ are the coincidence counts recorded between D$_{T}$ and D$_{1}$ (D$_{2}$). Each coincidence count was recorded over a $10$ s integration time. Also shown in Fig.~\ref{AvsG} as colored areas is an error model based on Monte Carlo simulations. This model aims to account for experimental imperfections that may affect our results \cite{SupMat}.

The solid, dashed, and dotted lines represent the theoretical predictions. We can observe that, depending on the values of $f$, there is an increase in the corresponding $\gamma_{c}$, indicating that the protocol becomes more robust against ADC when stochastic filters are applied. Consider the case without any filtering ($f=0$), the DW decreases rapidly as a function of $\gamma$, falling below the classical limit (represented by the horizontal line) when $\gamma_c \approx 0.38$. However, when $f=0.45$ ($f=0.90$) the DW can beat the classical bound until $\gamma_{c}=0.52$ ($\gamma_{c}=0.82$), beyond which the scheme no longer improves.  The red dots in the green region also decrease rapidly as a function of $\gamma$. However, there is an improvement when $F_{A}$ and $F_{B}$ are applied with $f_{a}=0.44\pm0.05$ and $f_{b}=0.44\pm0.06$ ($f_{a}=0.88\pm0.01$ and $f_{b}=0.91\pm0.01$), shown in the red (yellow) region. Therefore, the experimental data show a good agreement with the theoretical predictions and demonstrate the activation of the SDI QRAC scheme for values of $\gamma \ge \gamma_c$ when filtering operations are applied.

\textit{Conclusion.}-- We have presented a $2^2 \xrightarrow {\text{p}} 1$ QRAC protocol and examined its performance under an ADC. To estimate the effect of an ADC, the maximum tolerable noise beyond which the ASP of the QRAC protocol is entirely degraded is quantified. In order to avoid such deterioration, a scheme for activating QRACs by employing suitable stochastic operations, resulting in the revival of the ASP is proposed.  Naturally, this method extends to any SDI scenario when testing a suitable DW for certifying revival of non-classicality against any hidden variable model (e.g. causal models \cite{KA24}).

Finally, given the interest in QRAC as a primitive in quantum information protocols, we expect our results to find use in other applications, such as random number generation, noise-adaption for self testing of untrusted devices to name a few. Furthermore, we hope that our work motivates future experimental investigations for noise-adapting higher dimensional QRACs and to explore activation protocols for other canonical noise channels. 

\begin{acknowledgements}
\emph{Acknowledgments.}---HSK and ASH thank A R Usha Devi, R Srikanth, and M \.{Z}ukowski for the discussions. HSK and MP thank NCN Poland, ChistEra-2023/05/Y/ST2/00005 under the project Modern Device Independent Cryptography (MoDIC). This work was initiated when ASH and MP were supported by the Foundation for Polish Science (IRAP project, ICTQT, contract No. 2018/MAB/5, co-financed by EU within Smart Growth Operational Programme, and HSK (and MP) was supported by the NCN through Grant No. SHENG (2018/30/Q/ST2/00625). FMQ acknowledges financial support from ANID PFCHA/MAGISTER NACIONAL. SG, SPW, GL, and ESG acknowledge support from the Fondo Nacional de Desarrollo Científico y Tecnológico (FONDECYT) (Grant Nos. 3210359, 1240746, 1200859, and 1231940), and the National Agency for Research and Development (ANID) Millennium Science Initiative Program—ICN17-012.
\end{acknowledgements}

\bibliographystyle{apsrev4-2}
\bibliography{aqrac.bib}

\end{document}


\title{Supplemental information for ``Noise-Resilient Quantum Random Access Codes''}
\author{H. S. Karthik$^{\ddagger}$}
\email{hsk1729@gmail.com}
\affiliation{International Centre for Theory of Quantum Technologies, University of Gda{\'n}sk, 80-308 Gda\'nsk, Poland.}

\author{S. G\'omez$^{\ddagger}$}
\affiliation{Departamento de Física, Universidad de Concepción, Casilla 160‑C, Concepción, Chile.}
\affiliation{Millennium Institute for Research in Optics, Universidad de Concepción, Casilla 160‑C, Concepción, Chile.}

\author{F. M. Quinteros}
\affiliation{Departamento de Física, Universidad de Concepción, Casilla 160‑C, Concepción, Chile.}
\affiliation{Millennium Institute for Research in Optics, Universidad de Concepción, Casilla 160‑C, Concepción, Chile.}

\author{Akshata Shenoy H.}
\affiliation{International Centre for Theory of Quantum Technologies, University of Gda{\'n}sk, 80-308 Gda\'nsk, Poland.}

\author{M. Paw{\l}owski}
\affiliation{International Centre for Theory of Quantum Technologies, University of Gda{\'n}sk, 80-308 Gda\'nsk, Poland.}

\author{S. P. Walborn}
\affiliation{Departamento de Física, Universidad de Concepción, Casilla 160‑C, Concepción, Chile.}
\affiliation{Millennium Institute for Research in Optics, Universidad de Concepción, Casilla 160‑C, Concepción, Chile.}

\author{G. Lima}
\affiliation{Departamento de Física, Universidad de Concepción, Casilla 160‑C, Concepción, Chile.}
\affiliation{Millennium Institute for Research in Optics, Universidad de Concepción, Casilla 160‑C, Concepción, Chile.}

\author{E. S. G\'omez}
\email{estesepulveda@udec.cl}
\affiliation{Departamento de Física, Universidad de Concepción, Casilla 160‑C, Concepción, Chile.}
\affiliation{Millennium Institute for Research in Optics, Universidad de Concepción, Casilla 160‑C, Concepción, Chile.}
\maketitle

\section*{A: Heralded photon source}
A continuous-wave laser at $405$ nm pumps a $20$ mm long periodically poled potassium titanyl phosphate (PPKTP - type-II) crystal, producing a pair of photons with orthogonal polarization at $810$ nm. The pump laser power is controlled actively (@$2$ mW) to reduce intensity fluctuations. A high-quality narrow bandpass filter with $0.5$ nm bandwidth (peak transmission $>95\%$) ensures frequency degeneracy. Moreover, single-mode optical fibers are used to filter the undesirable spatial correlations of the down-converted photons. A numerical model is considered in the setup design to maximize the coincidences rate \cite{PhysRevA.72.062301}. The optimal coupling condition is reached when $\omega_{DCP}=\sqrt{2}\omega_{p}$, where $\omega_{p}$ and $\omega_{DCP}$ are the waist modes of the pump beam and the down-converted photons at the center of the PPKTP crystal, respectively. Our configuration achieves these waists with a lens $L$ of $20$ cm focal length at the front of the crystal and $10\times$ objective lenses for coupling the down-converted photons. The coincidence counts between the trigger detector D$_{T}$ and Bob's detectors D$_{1}$ (D$_{2}$), announce the presence of the heralded single-photon event used in the experiment. In addition, a manual fiber polarization controller (MPC) is used to adjust the initial polarization of Alice's photon.
\renewcommand{\theequation}{B-\arabic{equation}}
\setcounter{equation}{0}

\section*{B: Sagnac interferometer as an amplitude damping channel}
As demonstrated and described in previous work \cite{almeida07,salles08,farias12,aguilar14}, an interferometer can be used to implement a decoherence channel. Here, we briefly describe the two-path Sagnac interferometer used to implement the ADC in the experiment, illustrated in Fig.(2) of the main text. The photon is first incident onto a polarizing beam-splitter (PBS), which separates the horizontal and vertical components of the polarization qubit into two linear momentum modes $\ket{0}$ and $\ket{1}$, depicted as continuous and dashed lines in Fig.{2} of the main text, respectively. The PBS can be thought of as a controlled-NOT (CNOT) gate between the polarization and momentum degrees of freedom, described as
\begin{equation*}
    \text{PBS} = \hat{h}\otimes(\ket{0}\bra{0} + \ket{1}\bra{1}) + i\hat{v}\otimes(\ket{0}\bra{1} + \ket{1}\bra{0}),
\label{eq:PBS}
\end{equation*}
where $\hat{h}=\ket{H}\bra{H}$ and $\hat{v}=\ket{V}\bra{V}$. Applying the PBS to an input state $\ket{\Psi}=\ket{\psi}\ket{0}$, where $\ket{\psi}$ is the initial polarization state, we obtain the following.
\begin{align*}
    |\Psi^{'}\rangle =\hat{h}\ket{\psi}\ket{0} + i \hat{v}\ket{\psi}\ket{1}.
\end{align*}
Inside the interferometer, a half-wave plate HWP$(\theta_{1})$ is added on the path $1$ with its fast axis aligned at angle $\theta_{1}$. Experimentally, it is important to consider the optical path length difference between the paths $0$ and $1$ that is caused by HWP$(\theta_{1})$. Hence, another half-wave plate is necessary in the path $0$ to compensate for the length of the optical path. After the wave plate, the state is described as 
\begin{align*}
    |\Psi^{''}\rangle =& \hat{h}\ket{\psi}\ket{0} + i \hat{\text{H}}(\theta_{1})\hat{v}\ket{\psi}\ket{1},
\end{align*}
where $\hat{\text{H}}$ is the Jones matrix representation of a half-wave plate \cite{saleh2019fundamentals}. The two paths are routed using the three mirrors so that the photon passes through the same PBS again,  implementing a second CNOT gate as described in \eqref{eq:PBS}. After the PBS, we have the following state:
\begin{equation}
    |\Psi^{'''}\rangle = \Big[\hat{h} - \hat{v}\hat{\text{H}}(\theta_{1})\hat{v}\Big]\ket{\psi}\ket{0} + i\Big[\hat{h}\hat{\text{H}}(\theta_{1})\hat{v}\Big]\ket{\psi}\ket{1}.
\end{equation}
The two output modes of the PBS are then combined incoherently at a beam splitter (BS), such that the optical path lengths are different by a few millimeters, which is much more than the longitudinal coherence length of the photons (a few hundred microns).  This is the experimental equivalent of ``tracing out" the environment, which here is the linear momentum degree of freedom. The output state is then described by a density matrix, 
\begin{equation}
    \rho_{out} = K_0 \ket{\psi} \bra{\psi} K_0^\dagger + K_1\ket{\psi} \bra{\psi} K_1^\dagger
\end{equation}
where the Kraus operators are given by
\begin{align*}
    K_0 =&  \hat{h} - \hat{v}\hat{\text{H}}(\theta_{1})\hat{v},\\
    K_1 =& \hat{h}\hat{\text{H}}(\theta_{1})\hat{v}. 
\end{align*}
Replacing the components of the Jones matrix makes it possible to write these Kraus operators in the form given in Eq. (13) of the main text.

\section*{C: Stochastic operations}
\begin{figure}[t!]
\includegraphics[scale =0.35]{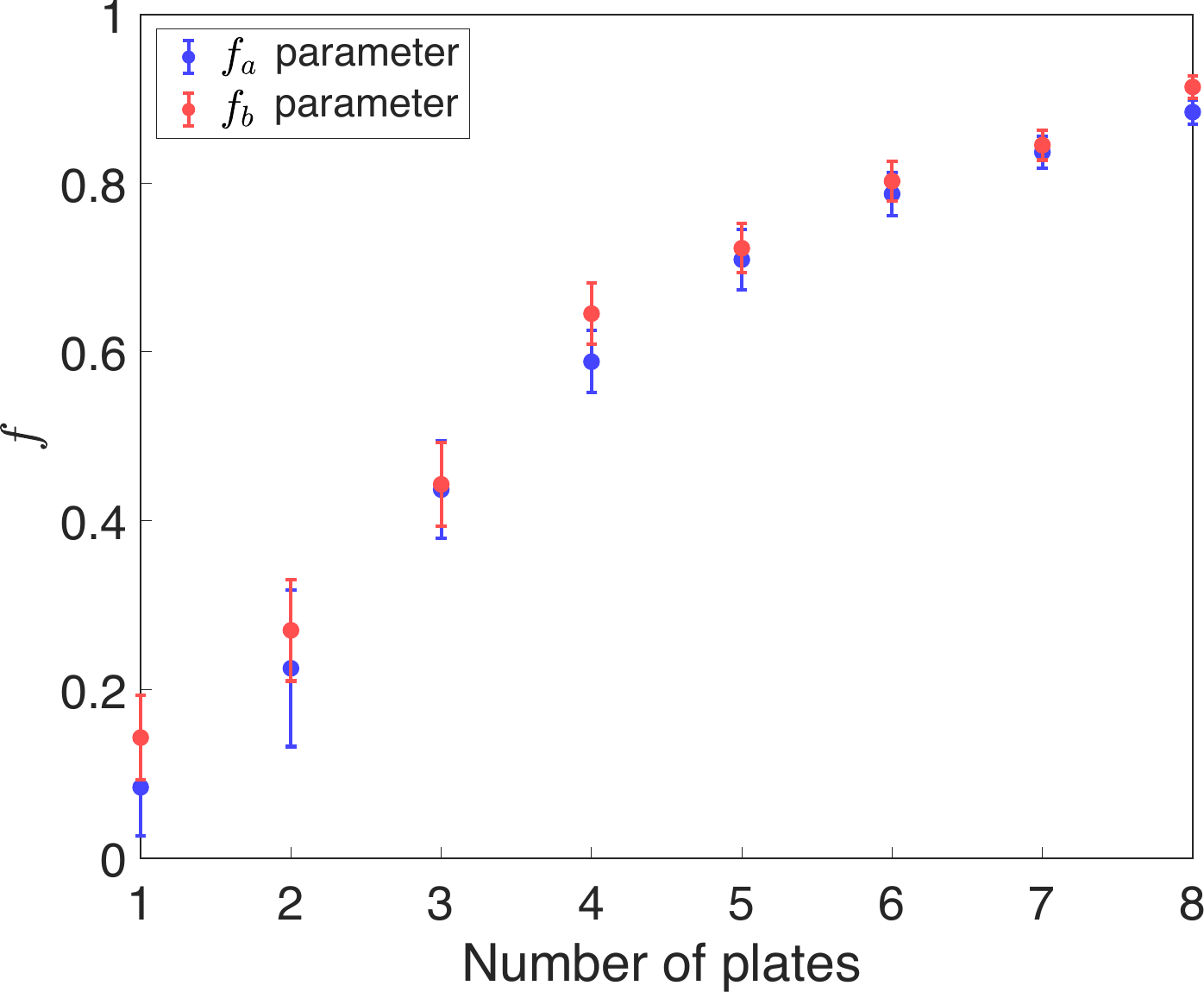}
\caption{Filter parameters $f_{a}$ and $f_{b}$ as a function of the number of glasses. The blue and red points represent the experimental results. These results were obtained using down-converted photons (@$810$ nm), with error bars calculated using Gaussian error propagation and Poisson statistics of the coincidence counts.}
\label{Filtro}
\end{figure}

 Figure \ref{Filtro} shows the corresponding parameter $f$ for different amounts of glass plates. We denote the $f_{a}$ ($f_{b}$) parameter when filtering the vertical (horizontal) polarization, represented by the blue (red) dots. To illustrate the advantages of using stochastic filters in this protocol, we consider two sets of glass configurations: $3$ and $8$ glass plates, corresponding to filter parameters of $f_{a} \approx \{0.44\pm0.05, 0.88\pm0.01\}$ and $f_{b} \approx \{0.44\pm0.06, 0.91\pm0.01\}$. 

\section*{D: Error analysis}
An error model based on Monte Carlo simulations is developed to show how experimental imperfections can affect our results. To simulate the error from Alice's state preparation, a mismatch of $\pm 1^{o}$ on the angle of the HWP$(\theta_{a})$ is considered. In addition, in the simulation, a deviation of $\pm1\%$ in the characterization of $f_{a}$ and $f_{b}$ is considered. In addition, a mismatch of $\pm 1^{o}$ on the HWP$(\theta_{b})$ placed in the measurement has been considered.

\bibliography{aqrac.bib}